\documentclass[a4paper,10pt]{article}
\usepackage{amssymb,graphicx}
\begin{document}

\begin{center}
{\Large\bf Supersymmetry Breaking and Gravitino Production \\ after Inflation in Modular Invariant Supergravity} \\
\vspace{16pt}
Kenji Takagi$^{b}$, Yuta Koshimizu$^{a}$, Toyokazu Fukuoka$^{a}$, Hikoya Kasari$^{a}$\\
 and Mitsuo J. Hayashi$^{a}$ \\
\vspace{16pt}
$^{a}${\it Department of Physics, Tokai University, 1117, Kitakaname, Hiratsuka, 259-1292, Japan} \\
$^{b}${\it Company VSN, Minato, Shibaura, 108-0023, Japan} \\

\vskip 16pt
E-mail: mhayashi@keyaki.cc.u-tokai.ac.jp
\vskip 16pt
\today
\end{center}

\thispagestyle{empty}
\vspace{24pt}

\begin{abstract}
By using a string-inspired modular invariant supergravity, which was proved well to explain WMAP observations appropriately, a mechanism of supersymmetry breaking (SSB) and Gravitino Production just after the end of inflation are investigated. 
Supersymmetry is broken mainly by F-term of the inflaton superfield and the Goldstino is identified to be inflatino in this model, which fact is shown numerically. By using the canonically normalized and diagonalized scalars, the decay rates of these fields are calculated, for both the $T$ and $Y$ into gravitinos. 
Non-thermal production of gravitinos is not generated from the inflaton (dilaton), since the inflaton mass is lighter than gravitino, but they are produced by the decay of modular field $T$ and scalar field $Y$.
Because the reheating temperature $T_R$ is about order $\sim O(10^{10})$ GeV and the mass of gravitino is  $3.16 \times 10^{12}$ GeV, it is not reproduced after the reheating of the universe. The gravitinos are produced almost instantly just after the end of inflation through $Y$ and $T$, not from inflaton. Because the decay time appears very rapid, gravitinos disappear before the BBN stage of the universe. The effects of the lightest supersymmetric particles (LSP) produced by gravitinos may be important to investigate more carefully, if the LSP's are the candidate of dark matter. 
\end{abstract}

\newpage
\section{Introduction}
\addtocounter{page}{-1}
Following ``Seven-Year Wilkinson Microwave Anisotropy Probe (WMAP) Observations" \cite{ref:WMAP}, the theory of inflation are proved to be the most promising theory of the early universe before the big bang.

 As far as the 4$D$, $N=1$ supergravity can play an elementary role in the theory of  the space-time and the particles \cite{ref:SUGRA}, it can also be essential in the theory of the early universe as an effective field theory. 
Supergravity, however, has been confronting with the difficulties, such as the $\eta$-problem and the supersymmetry breaking (SSB) mechanism has been studied by many authors \cite{ref:Nilles, ref:NoScale, ref:Witten, ref:Cvetic}.
We have investigated to prevail over these difficulties in Refs.\cite{ref:Hayashi1, ref:Hayashi2} by using the string-inspired modular invariant supergravity induced from superstring \cite{ref:Ferrara}. We found that the interplay between the dilaton field $S$ and gauge-singlet scalar $Y$ could give rise to sufficient inflation. The model we used, cleared the $\eta$-problem and appeared to predict successfully the values of observations at inflation era.
It predicted for examples, the indices $n_{s* } \sim 0.951$ and
$\alpha_{s*} \sim -2.50 \times 10^{-4}$. The value of $n_{s*}$ is consistent with the recent observations; the best fit of seven-year WMAP data using the power law $\Lambda$CDM model is
$n_s \sim 0.963 \pm 0.014$ \cite{ref:WMAP}.
The estimation of the spectrum was as $\mathcal{P_R}_* \sim 2.36\times10^{-9}$, which
result matches the measurements as well \cite{ref:WMAP, ref:Hayashi1, ref:Hayashi2}.  
 
In supergravity, gravitino is a unique object and cosmological meanings of gravitino is one of the most important problem as well as SSB mechanism. 
In this letter we will concentrate on the mechanism of SSB and the gravitino production just after the end of inflation. 

First we will briefly review the model and the former results \cite{ref:Hayashi1, ref:Hayashi2} as follows.
It is convenient to introduce the dilaton field $S$, a chiral superfield $Y$ and the modular field $T$. Here, all the matter fields are set to zero for simplicity.
Then, the effective No-Scale type K{\"a}hler potential and the effective superpotential that incorporate modular invariance are given by:
\begin{eqnarray}
&&K=-\ln \left(S+S^\ast\right)-3\ln \left(T+T^\ast-|Y|^2\right), \\
&&W=3bY^3\ln\left[c\>e^{S/3b}\>Y\eta^2(T)\right],
\end{eqnarray}
where $\eta(T)$ is the Dedekind $\eta$-function, defined by:
\begin{equation}
\eta(T)=e^{-2\pi T/24} \prod^{\infty}_{n=1}(1-e^{-2\pi nT}).
\end{equation}
The parameter $b$ and $c$ are treated as free parameters in this letter as discussed in Ref.\cite{ref:Hayashi2}.
The scalar potential is in order:
\begin{eqnarray}
&&V(S,T,Y)=\frac{3(S+S^\ast)|Y|^4}{(T+T^\ast-|Y|^2)^2}
\Bigg(3b^2 \left|1+3\ln\left[c\>e^{S/3b}\>Y\eta^2(T)\right]\right|^2
\nonumber\\
&&\qquad\qquad\quad+\frac{|Y|^2}{T+T^\ast-|Y|^2}
\left|S+S^\ast-3b\ln\left[c\>e^{S/3b}\>Y\eta^2(T)\right]\right|^2
\nonumber\\
&&\qquad\qquad\quad+6b^2|Y|^2\left[2(T+T^\ast)\left|\frac{\eta^\prime(T)}{\eta(T)}\right|^2
+\frac{\eta^\prime(T)}{\eta(T)}
+\left(\frac{\eta^\prime(T)}{\eta(T)}\right)^\ast\right]\Bigg),
\end{eqnarray}
where the potential is corresponding to canonically normalized kinetic Lagrangian. The potential is explicitly modular invariant and can be shown to be stationary at the self-dual points $T=1$ and $T=e^{i\pi /6}$ \cite{ref:Ferrara}, (see also \cite{ref:Nilles0}).

We had found that the potential $V(S,Y)$ at $T=1$ has a stable minimum at for the values
$b = 9.4$, $c = 131$ and obtained
\begin{eqnarray}
S_{{\rm min}} = 1.51, \qquad Y_{{\rm min}} = 0.00878480,
\end{eqnarray}
where $\eta (1) = 0.768225$, $\eta^2 (1) = 0.590170$, $\eta' (1) = -0.192056$, $\eta'' (1) = -0.00925929$ are used.
The inflationary trajectory can be well approximated by
\begin{equation}
Y_{\rm min}(S) \sim 0.009268 e^{-0.035461 S},
\end{equation}
which corresponds to the trajectory of the stable minimum for both $S$ and $Y$.

\section{Gravitino mass, Evolution of inflaton and F-term SSB}

Now we will briefly investigate the properties of inflaton $S$, gravitino and SSB mechanism. 
First, gravitino mass is given in this case 
\begin{equation}
m_{3/2} = M_P e^{K/2} |W| = 3.16 \times 10^{12} \,\, {\rm GeV},
\end{equation}
where $\hbar = 6.58211915 \times 10^{-25}$ GeV$\cdot$sec and $M_p = 2.435327 \times 10^{18}$ GeV are used.

If the chaotic potential
\begin{equation}
V = \frac{1}{2} m^2_\phi \phi^2, \label{chao_pot}
\end{equation}
is considered, the equation of motion of the scalar field is given by
\begin{equation}
\ddot{\phi} + \frac{2}{t} \dot{\phi} + m^2_\phi \phi = 0. \label{Reh_Sca_EOM}
\end{equation}

Then the general solution of this differential equation is obtained as
\begin{equation}
\phi (t) = \frac{c_1}{m_\phi t} \sin ( m_\phi t) + \frac{c_2}{m_\phi t} \cos ( m_\phi t). \label{Reh_Gen_Sol}
\end{equation}
Here, by taking limit $t \to 0$, $c_2 = 0$ follows, and if the amplitude $\bar{\phi}(t)$ is defined 
\begin{equation}
\bar{\phi} (t) \equiv \frac{c_1}{m_\phi t},
\end{equation}
then solution is damping oscillation
\begin{equation}
\phi (t) = \bar{\phi} (t) \sin ( m_\phi t) \label{Reh_Gen_Sol2}.
\end{equation}
In our model, by expanding $V$ around the minimum of $S(t)$, $Y(t)$ and
 fixed $T=1$, and by providing $S(t)$ and $Y(t)$ are real, then we obtained $S(t)$, $Y(t)$ as follows:
\begin{eqnarray}
&&S(t) = S_{{\rm min}}+\sqrt{\frac{8}{3}}\frac{\sin(m_S t)}{m_S t}, \\
&&Y(t) = \frac{1}{\eta^2 (1) e^{1/3} c} e^{-\frac{S(t)}{3b}}.
\end{eqnarray}

In order to argue on the evolution of $F-$terms, $m^i$ is defined as
\begin{eqnarray}
m^i \equiv {\cal D}^i m = e^{K/2} \left[ \partial^i W + ( \partial^i K ) W \right],
\end{eqnarray}
where $m \equiv e^{K/2} W$.
The $F-$term SSB scale is given by Nilles et al. \cite{ref:Nilles1, ref:Nilles2, ref:Nilles3}
\begin{eqnarray}
{f_{\phi_i}}^2\equiv {m_i}^2+\frac{1}{2}\left(\frac{d\phi_i}{dt} \right)^2,
\end{eqnarray}
where ${f_{\phi_i}}^2$ give a ``measure'' of the size of the SSB provided by the $F-$term associated with the $i$-th scalar field, which is the same as $\alpha$ in Kallosh et al. \cite{ref:Kallosh}. The factor $\frac{1}{2}$ in front of $\dot{\phi_i}$ shows $\dot{\phi_i}$ are real. 

In our model these quantities are estimated by calculating  
$m^SG^{-1}{}^{S}_{S} m_{S}$, $m^YG^{-1}{}^{Y}_{Y} m_{Y}$, $m^TG^{-1}{}^{T}_{T} m_{T}$ and time derivatives of these fields, 
finally ${f_{\phi_i}}^2$ are given by
\begin{eqnarray}
f^2_{S'} = 1.43 \times 10^{25}, \qquad f^2_{Y'} = 1.24 \times 10^{21}, \qquad f^2_{T'} = 4.56 \times 10^{23},
\end{eqnarray}
where numerical values are estimated at the stationary points $S_{{\rm min}}$, $Y_{{\rm min}}$ which are also asymptotic values.
$S'$, $Y'$, $T'$ are mass eigenstates that are canonically normalized \cite{ref:Endo}, \cite{ref:nakamura} as follows: 
\begin{eqnarray}
&&S' = 3.00 \times 10^{-1} S + 1.94 \times 10^{-3} Y - 3.66 \times 10^{-1} T \\[5pt]
&&Y' = 3.82 \times 10^{-4} S + 1.22 \, Y - 2.94 \times 10^{-8} T \\[5pt]
&&T' = 1.40 \times 10^{-1} S - 7.49 \times 10^{-3} Y + 7.85 \times 10^{-1} T.
\end{eqnarray}

The evolution of the ratios is shown at Fig.1
\begin{figure}[!htbp]
\begin{center}
\includegraphics[width=7cm]{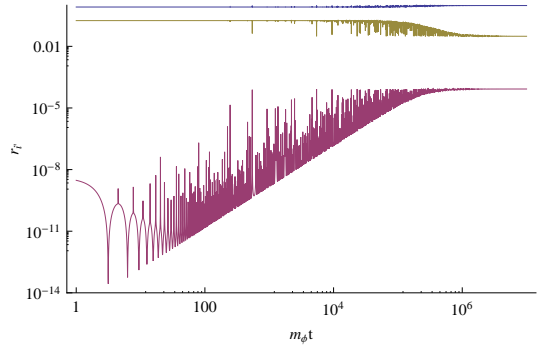}
\caption{Relative contribution of the three scalar fields $S'$, $Y'$ and $T'$ to the supersymmetry breaking during their evolution.
That of $r_{S'}$ corresponds to the highest curve, $r_{T'}$ the middle, $r_{Y'}$ the lowest.
}
\label{fig:one}
\end{center}
\end{figure}

By inserting the stationary values of $S'$, $Y'$, $T'$, we obtain the ratios $r_{S'}$, $r_{Y'}$, $r_{T'}$
\begin{eqnarray}
&&r_{S'} =\frac{f^2_{S'}}{f^2_{S'}+f^2_{Y'}+f^2_{T'}}=0.969, \\[5pt]
&&r_{Y'} =\frac{f^2_{Y'}}{f^2_{S'}+f^2_{Y'}+f^2_{T'}}=8.38 \times 10^{-5}, \\[5pt]
&&r_{T'} =\frac{f^2_{T'}}{f^2_{S'}+f^2_{Y'}+f^2_{T'}}=3.09 \times 10^{-2}.
\end{eqnarray}

Thus, supersymmetry is overwhelmingly broken by superfield $S'$. Contrary to the important fact of the interchange of supersymmetry breaking fields pointed out by Nilles et al. \cite{ref:Nilles1, ref:Nilles2, ref:Nilles3}, it does not occur in our model.

The values of masses of supersymmetric partners $\tilde{S}$, $\tilde{Y}$, $\tilde{T}$
are obtained as follows.
Using $m^{ij} = m^{ji}$ and $\chi_i \chi_j = - \chi_j \chi_i \,\, (i \neq j)$, we obtain: 
\begin{eqnarray}
m^{ij} \chi_i \chi_j 
= m^{\tilde{S}\tilde{S}} \chi_{\tilde{S}} \chi_{\tilde{S}} 
+ m^{\tilde{Y}\tilde{Y}} \chi_{\tilde{Y}} \chi_{\tilde{Y}}
+ m^{\tilde{T}\tilde{T}} \chi_{\tilde{T}} \chi_{\tilde{T}}, 
\end{eqnarray}
where we have neglected the Hermite conjugate terms.
Canonically Normalized fermionic states are given by
\begin{eqnarray}
\tilde{S}' = 0.331 \tilde{S}, \qquad
\tilde{Y}' = 1.22 \tilde{Y}, \qquad
\tilde{T}' = 0.867 \tilde{T} - 7.61 \times 10^{-3} \tilde{Y}.
\end{eqnarray}
Then, the values are numerically determined as
\begin{eqnarray}
m_{\tilde{S}'} = 0 \,\, {\rm GeV}, \qquad
m_{\tilde{Y}'} = 3.01 \times 10^{17} \,\, {\rm GeV}, \qquad
m_{\tilde{T}'} = 2.65 \times 10^{15} \,\, {\rm GeV}.
\end{eqnarray}

Since $\tilde{S}$ is massless and $S$ breaks supersymmetry, $\tilde{S}$ state is identified with Goldstino, which is absorbed into gravitino by super-Higgs mechanism \cite{ref:SUGRA, ref:moroi}.

\section{Gravitino productions from heavy scalar bosons}

Now we will investigate the gravitino production from heavy scalar bosons after inflation.
The interaction terms between scalar fields $\phi_i$ and gravitino $\psi_\mu$ in the total Lagrangian density of supergravity are selected as follows \cite{ref:moroi}:
\begin{eqnarray}
&&e^{-1}{\cal{L}}_{int}=\epsilon^{\mu\nu\rho\sigma}\bar{\psi}_\mu\bar{\sigma}_{\nu}\partial_\rho\psi_\sigma
+\frac{1}{4}\epsilon^{\mu\nu\rho\sigma}\bar{\psi}_\mu\bar{\sigma}_{\nu}\left(G_j\partial_\rho\phi^j-G_{j^*}\partial_\rho\phi^{*j} \right)\psi_\sigma\nonumber\\
&&\qquad\qquad\quad -e^{G/2}\left(\psi_\mu\sigma^{\mu\nu}\psi_\nu+\bar{\psi}_\mu\bar{\sigma}^{\mu\nu}\bar{\psi}_\nu\right).
\end{eqnarray}

These interaction terms are expanded by the shift $\delta \phi_i$ from each stable value for each $\phi_i$'s, i.e., $\delta\phi_i=\phi_i-\left<\phi_i\right>$.
In our model there are three scalar fields $S,Y,T$ corresponding to $\phi_i$'s.

$S,Y,T$ are canonically normalized by ${\phi_i}'=\sum_j\alpha^i_j\tilde{\phi^j}$, where 
the coefficients of canonical normalization $\alpha^i_j$ are defined in the later of this paper.
Since $\left<G_i\right>$'s are also affected by the normalization, the normalized $\left<{G_i}\right>$'s are replaced with $\left<G_i'\right>$s and 
$\alpha^i_j$'s are replaced by $\left<{\alpha^i_j}\right>$s at the stable points $\left<\tilde{\phi_i}\right>$.

By using these formula and general relation on gravitino mass $m_{3/2}=\left<e^{G/2}\right>$,
the interaction terms are obtained as:
\begin{eqnarray}
-\frac{1}{8\sqrt{2}}m_{3/2}\left<{G_i}' \right>\left<{\alpha^i_j}\right>\tilde{\phi^j}\bar{\psi}_\mu[\gamma^{\mu},\gamma^\nu]\psi_\nu.
\end{eqnarray}

The helicity $1/2$ part of massive gravitino is defined by the tensor product of vector and spinor as \cite{ref:Auvil} 
\begin{eqnarray}
u_\mu(k;1/2)\simeq i\sqrt{\frac{2}{3}}\epsilon_\mu(k;0) u(k;1/2)+i\sqrt{\frac{1}{3}}\epsilon_\mu(k;1)u(k;-1/2),
\end{eqnarray}
where the coefficient of each term is Clebsch-Gordan coefficient, $\epsilon_\mu(k;\lambda)$ is wave function of vector field with helicity $\lambda$ and $u(k;h)$ is spinor wave function with helicity $h$.

The decay rate $d\Gamma=\frac{|{\bf{k}}|}{8\pi m_\phi^2}|{\cal{M}}|^2\frac{d\Omega}{4\pi}$ is now in order \cite{ref:Endo, ref:nakamura},:
\begin{eqnarray}
\Gamma(\phi\rightarrow \psi_{3/2}+\psi_{3/2})=\frac{\left<{G_i}' \right>^2\left<{\alpha^i_j}\right>^2m_\phi^3}{288\pi}\left(1-\frac{2m^2_{3/2}}{m_\phi ^2}\right)^2\left(1-\frac{4m^2_{3/2}}{m_\phi ^2}\right)^\frac{1}{2}\label{decayrate2}.
\end{eqnarray}

After scalars $S,Y,T$ are canonically normalized and the masses diagonalized, the mass eigenstates are denoted by $S',Y',T'$, then masses are calculated as 
$M_{S'}=3.97 \times 10^{12}$ GeV, $M_{Y'}=2.45 \times 10^{17}$ GeV, $M_{T'}=9.02 \times 10^{12}$ GeV. 

Since it is impossible that $S'$ decay into the gravitino, Only two decay processes $Y'\rightarrow\psi_{3/2}+\psi_{3/2}$ and $T'\rightarrow\psi_{3/2}+\psi_{3/2}$ are enough to concern. By inserting canonical normalization factors and eigen mass values into (\ref{decayrate2}) that are independent on the features of canonical normalization and diagonalization, 
the decay rates and the decay times are obtained as follows
\begin{eqnarray}
&&\Gamma(Y' \rightarrow \psi_{3/2}+\psi_{3/2}) \simeq 3.42 \times 10^8 \quad\,\,\,\, {\rm GeV}, \\
&&\tau (Y' \rightarrow \psi_{3/2}+\psi_{3/2})  \simeq 1.93 \times 10^{-33} \quad\,\, {\rm sec}, \\
&&\Gamma(T' \rightarrow \psi_{3/2}+\psi_{3/2}) \simeq 2.55 \times 10^{-3} \quad {\rm GeV}, \\
&&\tau (T' \rightarrow \psi_{3/2}+\psi_{3/2})  \simeq 2.59 \times 10^{-22} \quad\,\, {\rm sec},
\end{eqnarray}
where the unit is changed from Planck unit $M_P=1$ to practical unit by dividing by $M_P^2$.
These processes occurs almost instantly. 

Because the primordial gravitinos decay very rapidly and the reheating temperature is lower than the gravitino mass,  the effect to the standard Big Bang Nucleosynthesis (BBN) scenario \cite{ref:takahashi, ref:kawasaki-moroi, ref:Kawasaki, ref:kawasaki2, ref:riotto, ref:Buchmuller}) may be negligible in our model (see also \cite{Weinberg:1982zq, Khlopov:1984pf}).

\section{Decay modes of heavy particles}

The decay modes of particles in this model are considered by using the interaction terms in supergravity Lagrangian density, which are as follows:
\begin{eqnarray}
{\mathcal{L}}_{\rm{interaction}}&=&e e^{\frac{K}{2}}\Big\{W^*\psi_\mu\sigma^{\mu\nu}\psi_\nu+W\bar{\psi}_\mu\bar{\sigma}^{\mu\nu}\bar{\psi}_\nu+\frac{i}{2}\sqrt{2}D_iW\chi^i\sigma^\mu\bar{\psi}_\mu \nonumber \\
&&\quad+\frac{i}{2}\sqrt{2}D_{i^*}W^*\bar{\chi}^i\bar{\sigma}^\mu\psi_\mu+\frac{1}{2}{\mathcal{D}}_iD_jW\chi^i\chi^j+\frac{1}{2}{\mathcal{D}}_{i^*}D_{j^*}W^*\bar{\chi}^i\bar{\chi}^j \nonumber \\
&&\quad-\frac{1}{4}g^{ij^*}D_{j^*}W^*f_{ab,i}\lambda^a\lambda^b-\frac{1}{4}g^{ij^*}D_iWf^*_{ab,j}\bar{\lambda}^a\bar{\lambda}^b\Big\} \nonumber \\
&&\quad-e e^{K}\big[g^{ij^*}D_iWD_{j*}W^*-3W^*W\big]\label{www}
\end{eqnarray}
where $K$ is K\"{a}hler potential, $W$ is superpotential, $\psi_\mu$ is gravitino,
$\chi^i$'s are fermionic superpartners corresponding to the scalar fields with indices $i$, $\lambda^a$ gauginos, $f_{ab}$ gauge kinetic function and so on $f_{ab,i}$'s mean the derivatives by scalar fields $\phi^i$ and finally the covariant derivative in these terms are defined by
\begin{eqnarray}
D_iW&=&W_i+K_iW\\
{\mathcal{D}}_iD_jW&=&W_{ij}+K_{ij}W+K_iD_jW+K_jD_iW-K_iK_jW-\Gamma^k_
{ij}D_kW
\end{eqnarray}
where $\Gamma^k_{ij}=g^{kl^*}g_{jl^*,i}$.

The interaction terms are obtained by expanding each term included in (\ref{www}) around the stable points of $S'$,$Y'$ and $T'$. From the first and the second term, the decay modes of the scalar fields $S'$,$Y'$ and $T'$ to gravitinos, provided that the modes satisfies the mass condition $m_{\phi^i}\ge 2 m_{\psi_\mu}$. From the third and the fourth terms,by replacing the gravitino field with Goldstiono $\psi$ which is the helicity $\pm\frac{1}{2}$ component of massive gravitino given by $\psi_\mu\simeq i\sqrt{\frac{2}{3}}\frac{1}{m_{3/2}}\partial_\mu\psi$.  In our model, since the Goldstino is identified with $\tilde{S'}$, the decays of $S',Y',T'\to \tilde{S'}+\tilde{S'},\tilde{S'}+\tilde{T'}$ and $\tilde{T'},\tilde{Y'}\to \tilde{S'}+S',\tilde{S'}+Y',\tilde{S'}+T'$are possibly occurs. From the fifth and sixth terms, pair productions $\tilde{S'},\tilde{Y'},\tilde{T'}$ from $S'$,$Y'$,$T'$, provided that the mass conditions $m_{\phi^i}\ge 2 m_{\tilde{\phi}^i}$ are satisfied. The seventh and eighth terms gives the decay modes from each scalar to gauginos, adding to the processes $S'\to\lambda^a+\lambda^a$, $Y'\to\lambda^a+\lambda^a$,$T'\to\lambda^a+\lambda^a$ will be possible. Finally, from the last term that defines the scalar potential, decay modes $Y'\to T'+T'$,$Y'\to S'+S'$,$Y'\to T'+S'$,$T'\to S'+S'$ can occur, after expanding the scalar potential around the stable points.
We show these varieties of decay modes at table \ref{Decay_modes}.
\\
\begin{table}[h]
\begin{center}
\begin{tabular}{|c|l|}
\hline
Parent particle & Decay modes \\
\hline
$S'$ & $\lambda^a+\lambda^a,$ other low energy scale particles \\
$Y'$ & $ S'+S',\tilde{S}'+\tilde{S}',\psi_\mu+\psi_\mu,\lambda^a+\lambda^a,T'+T',\tilde{T}'+\tilde{T}',S'+T',\tilde{S}'+\tilde{T}'$ \\
$T'$ & $ S'+S',\tilde{S}'+\tilde{S}',\psi_\mu+\psi_\mu,\lambda^a+\lambda^a$ \\
$\tilde{S}'$ & Goldstino, absorbed by gravitino $\psi_{3/2}$ making massive  \\
$\tilde{Y}'$ & $ S'+\tilde{S}',T'+\tilde{S}',Y'+\tilde{S}',S'+\tilde{T}',T'+\tilde{T}'$ \\
$\tilde{T}'$ & $ S'+\tilde{S}',T'+\tilde{S}'$ \\
$\psi_{3/2}$ & other low energy scale particles \\
\hline
\end{tabular}
\caption{Decay modes of fields in the model}
\label{Decay_modes}
\end{center}
\end{table}
$\,$ \\
The masses are obtained by our model setting as follows:
\begin{eqnarray*}
M_{S'}&=&3.97\times 10^{12} \,\,{\rm GeV}, \qquad\, M_{\tilde{S}'}=0 \,\,{\rm GeV} \\
M_{Y'}&=&2.45\times 10^{17} \,\,{\rm GeV}, \qquad M_{\tilde{Y}'}=3.01\times 10^{17} \,\,{\rm GeV} \\
M_{T'}&=&9.02\times 10^{12} \,\,{\rm GeV}, \qquad M_{\tilde{T}'}=2.65\times 10^{15} \,\,{\rm GeV} \\
m_{3/2}&=&3.16\times10^{12} \,\,{\rm GeV}
\end{eqnarray*}

Let us consider the cases of decay modes of $Y'$, as an 
example, $Y',\tilde{Y}'$ once decay into $T',\tilde{T}'$ and others,
further the secondary decay into lighter particles such as $S'$ and  $\psi_\mu$ (also possibly exist the modes directly from $Y'$into them), finally will decay into the Lightest SUSY particles (LSP) or, the Next to Lightest SUSY particles (NLSP) and ordinary standard theory particles. The problems here arise are the effects on the Big Bang Nucleosynthesis (BBN), which may give rise destruction of nuclei produced by BBN. And also, provided to identify the LSP are the candidates of the dark matter, the abundance might exceed the observational amount of dark matter.
It is the problem to calculate the yield variables of the LSP's. 
There seems, however, exist two problems to calculate the yield variables of LSP. Because the decays mainly happen before the equilibrium state of the universe, namely, they are produced in nonthermal processes, it makes  difficult to use the ordinary thermodynamical treatments. On the other hand, because the abundance of the MSSM, NMSSM particles depends on the contributions of $S'$ and $T'$ decays and direct decay of $Y'$, $Y'\rightarrow S'\rightarrow$ MSSM, NMSSM particles, it is quite complicate to analyse the amount of LSP particles. We will tackle with these points in our forthcoming paper.\cite{ref:preparation}

\section{Conclusion}

The model we used, cleared the $\eta$-problem and appeared to predict successfully the values of observations at inflation era.
It predicted for examples, the indices $n_{s* } \sim 0.951$ and
$\alpha_{s*} \sim -2.50 \times 10^{-4}$. The value of $n_{s*}$ is consistent with the recent observations; the best fit of seven-year WMAP data using the power law $\Lambda$CDM model is
$n_s \sim 0.963 \pm 0.014$ \cite{ref:WMAP}.
The estimation of the spectrum was as $\mathcal{P_R}_* \sim 2.36\times10^{-9}$, which
result matches the measurements as well \cite{ref:WMAP, ref:Hayashi1, ref:Hayashi2}.  
We have investigated on the preheating mechanism just after the end of inflation through both the inflaton (dilaton) decay into MSSM gauge sector and the collision of two inflaton into two righthanded sneutrinos. 
We have concluded in our former paper \cite{ref:Koshimizu} that the contribution of both the inflaton decays and the parametric resonance of four body scattering process play equally important roles in the preheating process just after the end of inflation. 
The reheating temperature is estimated is about order $\sim O(10^{10})$ GeV.

Because the mass of gravitino is calculated as $3.16 \times 10^{12}$GeV, it is rather heavy and may be unstable, therefore, may not be considered as LSP or NLSP and not a dark matter candidate discussed in Refs.\cite{ref:Endo, ref:nakamura, ref:Pradler}. However the main topic of supergravity at present stage of the theory is whether the gravitino exist or not in nature despite its mass.
It is not reproduced after the reheating of the universe. The gravitinos are produced almost instantly just after the end of inflation through $Y$ and $T$, not from inflaton. 
Because the reheating temperature $T_R$ is about order $\sim O(10^{10})$ GeV,  gravitinos are not reproduced after the reheating of the universe.
Because the decay time appears very rapid, gravitinos disappear before the BBN stage of the universe. The effects of LSP produced by gravitinos may be important to investigate more carefully, if the LSP's are the candidate of dark matter. 

The topic must be remained to later works \cite{ref:preparation}.
Therefore, we only remark here that our present model seems consistent with the present situation of observations.

On the other hand, supersymmetry is overwhelmingly broken by $F-$term of the inflaton (dilaton) superfield $S$, that may be contrary to the occurrence of the interchange of SSB fields pointed out in other type of models by Nilles et al. \cite{ref:Nilles1, ref:Nilles2, ref:Nilles3}. 

Though we have been exclusively restricted our attention to a model of Ref.\cite{ref:Ferrara}, the other models derived from the other type of compactification seems very interesting. Among them KKLT model \cite{ref:Linde, Endo:2005uy, Choi:2005uz, ref:Nilles4} attracts our interest, where the moduli superfield $T$ plays an essential roles. We should take all the circumstances into consideration on essential problems confronted in construction of string-inspired modular invariant Supergravity models.


\end{document}